\documentstyle[prb,aps,epsf,twocolumn]{revtex}

\begin{document}

\def\be{\begin{eqnarray}}
\def\ee{\end{eqnarray}}
\def\ts{\hskip 5pt}
\def\tts{\hskip 2pt}
\def\ospace{\o\hbox{ }}
\def\pt{\partial}

%\draft

\twocolumn[
\tightenlines

%ABSTRACT
\title{Flux Line Lattices in Artificially Layered Superconductors}

\author {A.M. Thompson and M.A. Moore}
\address{Theory Group, Department of Physics, University of
 Manchester,}
\address{Manchester, M13 9PL, U.K.}
\maketitle\widetext
\leftskip 54.8pt
\rightskip 54.8pt

\begin{abstract}
The flux line lattice of superconductors has been investigated when there
 exists a periodicity in the underlying system, such as can occur in
 artificially layered structures. 
For small fields parallel to the layers the flux lines enter the sample in
 sequential rows, with the possibility of jumps in the magnetization as new
 rows are created. 
As the field is increased these discontinuities gradually decrease, but there
 still exist transitions between states that are aligned differently to the
 periodic direction. 
Increasing the magnitude of the periodic potential reduces the competition
 between differently aligned lattices and tends to lock in one particular
 alignment. 
The effect of transitions on the shear modulus is also discussed and
 related to the experiments of Theunissen et al. 
\end{abstract}

\leftskip 54.8pt
\rightskip54.8pt

\pacs{PACS: 74.60.Ec, 74.60.Ge}
\narrowtext
]   

%
%%%%%%%%%%%%%%
%
 
\section{Introduction}

The nature of the flux line lattice in superconductors still attracts much
 interest. For an isotropic system the equilibrium lattice is known to be
 triangular. We will investigate situations where this may not be the case. 
In particular we shall impose a one-dimensional periodic structure on an
 infinite superconductor, and use this periodicity to model an underlying
 potential. 
This allows us to model for example some of the behavior of artificially
 layered structures when the field is applied parallel to the layers. 

The interplay between the flux line lattice and an underlying periodic
 potential has already been investigated for many different superconducting
 systems. 
These include thin films with modulated thickness\cite{martinolli}, Pb-Bi
 alloys with periodic concentration of Bi\cite{raffy}, Pb/Ge multilayers
 with a lattice of submicron holes\cite{metlushko}, superconducting wires
 with an hexagonal array of artificial pins\cite{cooley} and spatially
 modulated Josephson junctions\cite{oboznov}. 

Using London Theory, Brongersma et al.\cite{brongersma1} modeled the
 magnetization of thin Nb-Cu multilayers with the applied field almost
 parallel to the layers. The magnetic induction was allowed to change by
 permitting the size of the system to vary while keeping the number of flux
 lines fixed. 
As the applied field was increased, a series of maxima in the magnetization
 was observed. This was due to the flux lines reorganizing within the
 sample, where transitions occurred between states where the flux lines
 form rows that divide the sample into equal parts. 
This situation has also been investigated by numerically solving the
 time-dependent Ginzburg Landau equations coupled to Maxwell's equations
 in a homogeneous isotropic superconducting thin film\cite{bolech}. 
As the field applied parallel to the film was increased the magnetization
 also showed a series of maxima. 
Away from these maxima there were a series of discontinuities in the
 magnetization as more flux lines penetrated into the sample. 

Critical currents also show similar behavior. 
Discontinuities in the critical current have been observed at specific
 matching fields where the numbers of rows change.  
The positions of the maxima in the critical currents is sensitive to the
 system used, with YBCO, BSCCO and NbCu multilayers showing different
 field dependences\cite{ziese}.

\begin{figure}[h]
  \narrowtext
  \centerline{\epsfxsize=4.0cm \epsfbox{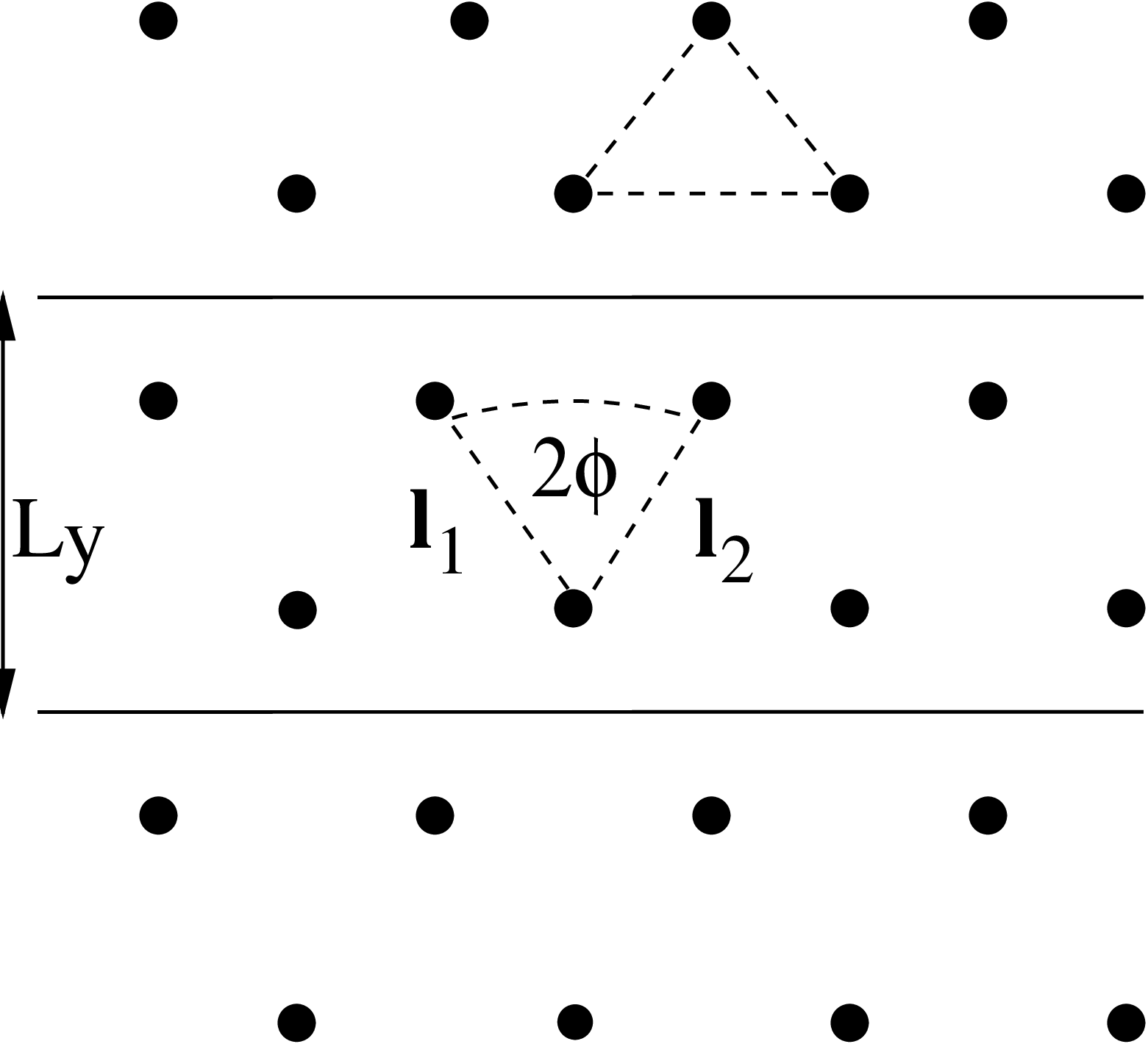}
  \qquad \epsfxsize=4.0cm \epsfbox{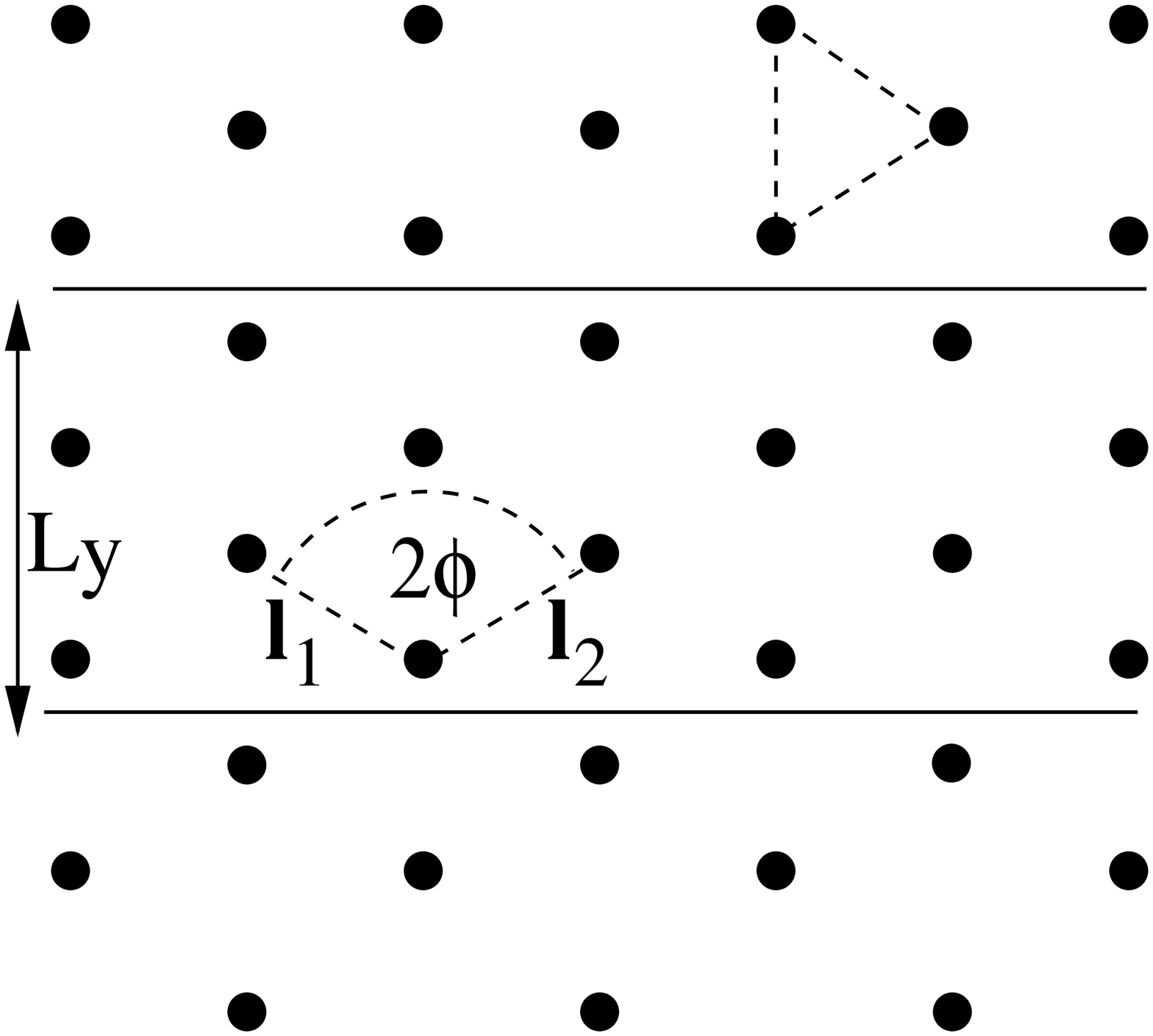} }
  \vglue 16.75pt
  \caption{A multilayered superconducting structure, with the field
 applied parallel to the layers. The filled circles represent the flux 
lines, which form a lattice commensurate with the principle region, a strip
 of width $L_y$. The lattices shown are centered rectangular structures,
 whose basis vectors are separated by the angle $2\phi$.
 The figure show a) $\phi = \pi/6$ and b) $\phi= \pi / 3 $ } 
  \label{fig:geom}
\end{figure}

We investigate here the properties of flux lines within an infinite system,
 under the influence of a potential with period $L_y$ in the $\hat{\bf y}$
 direction. 
We take the applied field to be in the $\hat{\bf z}$ direction. 
This infinite system contains a principal region of width $L_y$ and
 approximates a layered system. 
In the absence of the potential, it is well known that flux lines form a
 periodic lattice with hexagonal symmetry in an infinite system. 
With the periodic potential the flux lines may form a lattice that is
 commensurate with the principal region and we investigate rectangular and
 centered rectangular structures with this property. 
The rectangular structure has basis vectors in the $\hat{\bf x}$ and
 $\hat{\bf y}$ directions, and the centered rectangular structure, shown in
 Fig. \ref{fig:geom}, is characterized by the angle $2\phi$ between the
 basis vectors. 
For two different $\phi$ the centered rectangular lattice will be the
 triangular lattice expected in a pure infinite system. 
These correspond to $\phi = \pi / 6$ and $\phi = \pi/ 3$. 
Fig. \ref{fig:geom}a and Fig. \ref{fig:geom}b show $\phi=\pi/6$ and
 $\phi=\pi/3$ respectively. 
The state with $\phi=\pi/6$ has the base of its equilateral triangle
 aligned perpendicular to the periodic direction. Conversely, in the state
 with $\phi=\pi/3$ the base is aligned parallel to the periodic direction. 
We distinguish between these two states by referring to the centered
 rectangular structure with $\phi = \pi/6$ as the lattice aligned parallel
 to the periodic direction (the $\hat{\bf y}$ axis), and the structure with
 $\phi = \pi/3$ as the lattice aligned perpendicular to the periodic
 direction. 

On minimizing the Gibbs free energy it is found that the equilibrium
 lattice usually has $\phi$ only approximately $\pi/6$ or $\pi/3$, but the
 above distinction regarding alignments is used. 
The equilibrium lattice is often a competition between these two alignments,
 and this competition is most readily seen when the periodic potential is
 very weak. 

As the applied field is varied two distinct types of transitions are
 observed between some of these different structures. 
The simplest, Type A, transitions just involve the number of rows of flux
 lines within the principal region increasing by one, with no realignment
 of the lattice.  
Type B transitions occur between states aligned differently to the
 periodic direction, and during these transitions the number of rows
 changes significantly. 

The periodic potential can be chosen to favor having the flux lines close to
 either the edges or the center of the principal region. This ensures one of
 the alignments of the flux line lattice is favored over the other. 
The sequence of transitions therefore can be changed by increasing the
 strength of the underlying potential. 

The model used to investigate the flux line lattice is developed in
 Section \ref{sec:lll}. 
The formalism discussed concentrates on the lowest Landau level (LLL) limit
 in the mean field approximation. Throughout we shall neglect thermal
 fluctuations. The transitions are also investigated in the London limit. 
For many of these transitions notable jumps in the magnetization are
 observed. This is not the only property that can be effected by the
 periodicity of the system. 
Some of the consequences of the periodicity are developed in Section
 \ref{sec:pot}, where we investigate in particular the shear modulus
 $c_{66}$. It `oscillates' in a manner similar to that seen in
 experiments by Theunissen et al.\cite{kes}.

\section{Basic Formalism}
\label{sec:lll}

The experiments of Theunissen et al.\cite{kes} suggest the possibility of
 transitions between different configurations of flux lines both in the
 limit of low magnetic induction where London Theory will be valid, namely
 where $c_{66} \propto b$ and at higher inductions where
 $c_{66}\propto (1-b^2)$ when the LLL is more appropriate.
 We investigate the LLL first.

The Ginzburg-Landau free energy is 
\be
{\cal F}[\Psi] = \int d^3r \left[ \alpha(T) |\Psi(r)|^2 +
 {\beta\over 2}|\Psi(r)|^4 - {1 \over 2m} \Psi(r)^* D^2 \Psi(r)
 + {1 \over 2 \mu_0} B^2 \right],
\label{gl}
\ee
where $D^2 = {\bf D} \cdot {\bf D}$ and
 ${\bf D} = -i\hbar \nabla - 2e {\bf A}$, and the magnetic induction
 ${\bf B} = \nabla \times {\bf A}$. 
The LLL assumes the magnetic screening length is infinite. 
We will use the Landau gauge where ${\bf A} = B x \hat{\bf{y}}$, and
 ignore fluctuations in the magnetic field. 
In the LLL the order parameter is expanded in the set of eigenfunctions
 of $D^2$, using only the degenerate states which have the lowest eigenvalue. 
In the rectangular geometry these degenerate states are
\be
\Psi_p = \exp\left( -ipk_0 y  - { 1 \over 2 l_m^2 }
 (x - pk_0 l_m^2 )^2 \right),
\label{landaustates}
\ee
where $l_m$ is the magnetic length $l_m = (\Phi_0 / 2\pi B)^{1/2}$ and
 $k_0$ is the wavevector in the y-direction whose value is determined by
 the boundary conditions. 
The general LLL order parameter is
 $\Psi = Q \sum_{p=-\infty}^{\infty} c_p \Psi_p$, which is periodic in the
 y-direction over a length $L_y = 2\pi / k_0$. 
Obviously states which do not have this high degree of periodicity of their
 order parameter at mean field level could exist for a layered system.
 We shall ignore them as even with the imposed periodicity the phase
 diagram at mean field level is extremely rich.

%%figure: structures investigated.
\begin{figure}
  \narrowtext
  \centerline{\epsfxsize=7.5cm
  \epsfbox{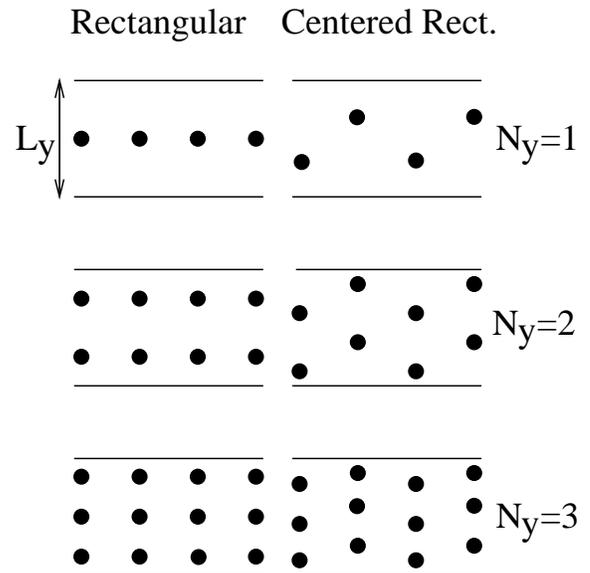} }
  \vglue 16.75pt
  \caption{The flux line lattices with rectangular and centered rectangular
 symmetry. These lattices are commensurate with the principal region, which
 is a strip of width $L_y$. The equilibrium lattice is predominantly a
 centered rectangular lattice.}
  \label{fig:struc}
\end{figure}

In an infinite system, the equilibrium configuration of flux lines is known
 to be that of a triangle\cite{kleiner}. 
In our problem the flux lines are imagined to interact weakly with an
 underlying periodic potential due to the layering. 
Although we ignore the direct contributions of it to the energy at this
 point, we assume that this potential acts to impose the periodicity $L_y$
 on the ordered state. 
Hence, the flux lines form a lattice that is commensurate with the principal
 region and repeats over a distance $l_y$, which must be a simple fraction
 of $L_y$, {\it i.e.} $l_y = L_y / N_y $ where $N_y$ is an integer equal
 to the number of rows of flux lines in the principal region. 
We investigate rectangular and centered rectangular structures with this
 property as shown in Fig. \ref{fig:struc}. The flux line lattice with
 rectangular symmetry has basis vectors ${\bf l}_1 = l_y \hat{\bf y}$ and
 ${\bf l}_2 = l_x \hat{\bf x}$. 
The centered rectangular structure, also shown in Fig. \ref{fig:geom}, has
 basis vectors ${\bf l}_1 = - l_x \hat{\bf x} + l_y \hat{\bf y}/2$ and
 ${\bf l}_2 = l_x \hat{\bf x} + l_y \hat{\bf y}/2$, and is characterized
 by the angle $2\phi$ between these two vectors. 
The model used is shown for a centered rectangular structure, but is
 easily modified for the rectangular lattice. 

Within the LLL the mean-field line $H_{c_2}$ is defined by $\alpha_B = 0$
 where $B=H_{c_2}$ and $\alpha_B = \alpha (T) + eB\hbar /m $. 
Using the linearized expression for $\alpha(T)$, 
$\alpha(T) = -\alpha^\prime(1 - t)$ then 
$\alpha_B = -\alpha^\prime (1 - t - b)$ where $t$ is the reduced temperature
 $T/T_{c0}$ and $b$ is the reduced field $B/B_{c0}$. 
$T_{c0}$ is the mean-field transition temperature while $B_{c0}$ is the
 straight line extrapolation of $H_{c_2}$ to zero temperature. 
The temperature is conveniently represented by the dimensionless parameter 
$\alpha_T = \alpha_B (\pi\hbar L_z / \beta e B k_BT )^{1/2}$, where
 $L_z$ is the sample thickness along the field direction. 
Low temperatures are represented by $\alpha_T \to -\infty$ while high
 temperatures correspond to $\alpha_T \to +\infty$. 

\begin{figure}
  \narrowtext
  \centerline{\epsfxsize=7.5cm
  \epsfbox{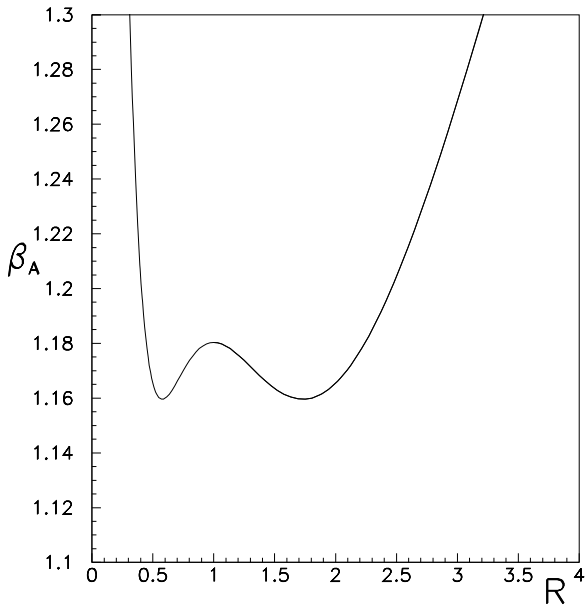} }
  \caption{The dependence of $\beta_A$ on the ratio $R=l_y/ 2 l_x$. 
    It should be noted that $\beta_A(R)= \beta_A(1/R)$ }
  \label{fig:beta}
\end{figure}

The rescaled free energy per flux line is then 
\be
{\cal F}_{flux}= -{k_BT\alpha_T^2 \over 2 \beta_A(R)}
 + {B^2 \over 2 \mu_0} A_{flux} L_z
\label{rescalefree} 
\ee
where $A_{flux} = \Phi_0 / B$ is the area per flux line.
 The ratio $R = l_y/2 l_x$, as used by Kleiner et al.\cite{kleiner},
 characterizes the lattice. 
Fig. \ref{fig:beta} shows the dependence of the Abrikosov parameter
 $\beta_A$ on the ratio $R$, and it should be noted that
 $\beta_A(R)= \beta_A(1/R)$,  where
 $\beta_A = \langle | \Psi |^4 \rangle / ( \langle | \Psi |^2 \rangle ^2 )$. 
The square lattice corresponds to $R=1$ and the minima in $\beta_A(R)$
 occur for the equilateral triangular lattice, with $R=1/\sqrt{3}$ and
 $R=\sqrt{3}$. These two minima correspond to the two alignments of the
 flux lattice relative to the periodic direction. The state with
 $\phi=\pi/6$ in Fig. \ref{fig:struc}a corresponds to the minima at
 $R=\sqrt{3}$ and the state with $\phi=\pi/3$ has $R=1/\sqrt{3}$.  

For a given field $B$, the free energy per unit volume $F_{vol}$ depends
 solely on the parameter $R$. 
If we minimize $F_{vol}$, the flux line lattice will be arranged in the
 configuration that minimizes $\beta_A(R)$, while keeping
 $B = 2RN_y^2 (\Phi_0/L_y^2)$.  
For small fields $B < \Phi_0/L_y^2$ the equilibrium configuration
 corresponds to $N_y = 1$ as the smallest value of $\beta_A(R)$ corresponds
 to $N_y=1$. 
As the field is increased there will be transitions between configurations
 of different $N_y$ and equal $\beta_A$. 
The transition between states $N_y =1$ and $N_y=2$ occurs when
 $B = 4 \Phi_0 / L_y^2$. 

For larger fields the transitions occur between states near the minima
 of $\beta_A$. 
Most transitions occur between states with $R \approx 1/\sqrt{3}$ but there
 are a few transitions to and from states near the minimum at $R=\sqrt{3}$. 
All these transitions occur for constant $B$ between states of
 equal $\beta_A(R)$ with a change in $N_y$.

%%%%
%%  Gibbs free energy
%%%%
\section{Gibbs free energy}
\label{sec:gibbs}

To understand the possibilities of transitions between different states at
 constant $H$ the relevant quantity to minimize is not the free energy per
 flux line, but rather the Gibbs free energy per unit volume ${\cal G}_{vol}$.
The Gibbs free energy per flux line can be written as
\be
{\cal G}_{flux} = - { \alpha_T^2 k_BT \over 2 \beta_A }
 + {1 \over 2 \mu_0} ( B - \mu_0 H )^2 L_ZA_{flux}
\label{gibbsflux}
\ee
Rewriting this in terms of the reduced fields Gibbs free energy per flux
 line becomes
\be
{\cal G}_{flux} = \frac12 \alpha_T^2 k_BT \left\{ 2 \kappa^2 \left( 
{ b - h \over 1- t -b } \right)^2 - {1 \over \beta_A} \right\}
\ee
where $h = \mu_0 H / B_{c0}$. 
Although $\alpha_T$ is the single intensive parameter that characterizes
 the free energy $\cal F$, it is necessary to show the explicit dependence
 on the average induction $B$. 
Therefore, since ${\cal G}_{vol} = {\cal G}_{flux} B / \Phi_0 L_Z $, at
 constant $H$ and $T$, 
\be
{\cal G}_{vol}(H,T,B) & = & \frac12 \mu_0 H_c^2  g(h,t,b) \nonumber \\
g(h,t,b) &=& 2 \kappa^2 (b - h)^2 - (1 - t -b)^2/\beta_A
\label{gibbsvol}
\ee
where $b$ is the reduced magnetic field $b=2 R N_y^2 (2\pi \xi^2 /L_y^2)$
 and $\xi$ is the coherence length defined through the upper critical field,
 $H_{c_2} = \sqrt{2}\kappa H_c = \Phi_0 / 2\pi \xi^2$. 
The behavior of the system is therefore controlled by $\kappa$ and
 $b_0 = 2 \pi \xi^2 /L_y^2$. 

%Small B limit
If the applied field $h$ is small, the Gibbs energy $g(h,t,b)$ is dominated
 by the term $(1 - t -b)^2/\beta_A$. 
The minimum free energy flux line configuration is obtained by being as
 close as possible to having $R=1/\sqrt{3}$ or $R=\sqrt{3}$. 
The range of allowed values of $b$ is small, and as $h$ is increased there
 is a series of first order transitions at which the lattice reconstructs. 
As the lattice undergoes a reconstruction between different states
 $\Psi(N_y,\sqrt{3} R)$ there are jumps in the magnetic induction $b$. 
Except for the transition $\Psi(1,1) \to \Psi(1,3)$ this involves a change
 in the number of `rows' of flux lines and these reconstructions can be
 classified into two categories. 
The simplest involve just an increase in the number of rows for two states
 with $R\approx 1/\sqrt{3}$ or $R\approx \sqrt{3}$. 
Some of these Type A transitions include $\Psi(2,1) \to \Psi(3,1)$ and
 $\Psi(4,1) \to \Psi(5,1)$ and occur between states with the same alignment
 to the periodic direction. 
The other set of transitions involve a change in the alignment of the flux
 line lattice. The Type B transitions occur between states near the different
 minima of $\beta_A(R)$ and involve either a {\it reduction} in the number
 of rows combined with a large increase in $R$, or an {\it increase} in the
 number of rows and a reduction in $R$. 
These transitions include $\Psi(3,1) \to \Psi(2,3)$ and
 $\Psi(4,3) \to \Psi(7,1)$. 
By putting in ascending order the values of $K N_y^2$ where $K=1/\sqrt{3}$
 or $\sqrt{3}$ we can see the appropriate sequence of states.
This sequence of states looks similar to that in the previous section, but
 the states possess a smaller range of values of $R$. 
Transitions occur between states with different $\beta_A$, different $b$ and
 different $N_y$.

%insert figure
\vglue -8.75pt
\begin{figure}
%  \narrowtext
  \centerline{\epsfxsize=6.5cm
  \epsfbox{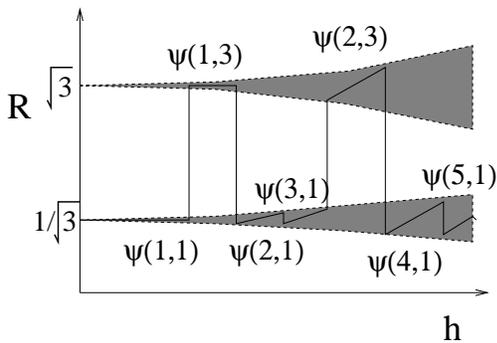} }
  \caption{A representation of how $R$ changes as $H$ increases. The range
   of values of $R$ that minimize the Gibbs free energy are shown within
   the shaded region.}
  \label{fig:scheme}
\end{figure}

%Large B limit
For very large applied fields, such that $2 \kappa^2(b - h)^2$ is the
 dominant term in $g(h,t,b)$, there is a wide range of equilibrium values
 of $R$ that are obtained as $h$ is varied. 
The lattice distorts from a triangular lattice to ensure $b \approx h$.
 However, transitions between states with different $N_y$ still occur, but
 these transitions involve very small changes in the magnetic induction $b$. 
%Schematic
As the applied field $h$ is increased there is a smooth crossover between
 these two limits as the two terms in $g(h,t,b)$ compete with each other. 
This results in the allowed values of $R$ spreading out from the two values
 of $1/\sqrt{3}$ and $\sqrt{3}$. 

These transitions can be shown schematically by looking at the allowed
 values of $R$ that minimize the Gibbs free energy. 
The transitions corresponding to small fields $h$ are between states defined
 by $R$ being very close to either $1/\sqrt{3}$ or $\sqrt{3}$. 
As $h$ increases, the range of values of $R$ that minimize the free energy
 increases, and this leads to the fractional jump in $b$ dropping.

%Jumps
%insert figure
\begin{figure}
%  \narrowtext
  \centerline{\epsfxsize=7.5cm
  \epsfbox{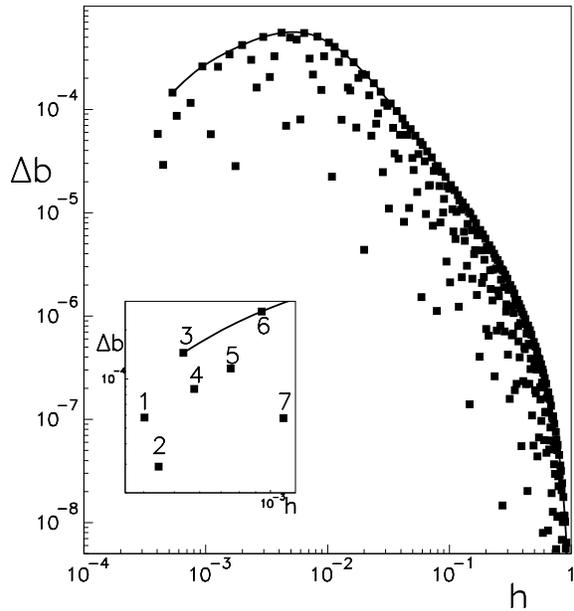} }
  \caption{The jumps in the magnetization vary as the applied field is
 increased, decaying to zero in the limit $h \to 1$; for $L_y = 10 \lambda$,
 $\kappa = 50$. The points indicate the change in the magnetization when
 the potential is very weak. The heavy line is a guide to the eye, whose
 physical significance is discussed in the next section. }
  \label{fig:dbvsh}
\end{figure}

The size of the jumps in the magnetic induction associated with these
 transitions is given in Fig. \ref{fig:dbvsh}. 
These jumps increase at small $h$ as approximately $h^{\frac12}$ but decay
 to zero in the limit $h \to 1$ following the heavy curve in
 Fig. \ref{fig:dbvsh}. 
The inset describes in detail the initial transitions, showing the points
 lying on or below the start of the heavy line. 
Within the LLL approximation the flux lines are not excluded from the sample
 at small $h$ as $H_{c_1} = 0$. 
The flux lines always form a centered rectangular structure and we only
 consider behavior for magnetic inductions larger than $B_{min}$.
 This is the magnetic induction that corresponds to a simple triangular
 lattice with $N_y = 1$ and $R=1/\sqrt{3}$. 
The first point in the inset corresponds to the transition between states
 $\Psi(1,1)$ and $\Psi(1,3)$. 
The subsequent sequence of these centered rectangular structures is
 $\Psi(2,1)$, $\Psi(3,1)$, $\Psi(2,3)$, $\Psi(4,1)$, $\Psi(5,1)$ and is
 easily predicted to the limit $h \to 1$. 
The Type A transitions, such as $\Psi(2,1) \to \Psi(3,1)$ (Point 3) and
 $\Psi(4,1) \to \Psi(5,1)$ (Point 6), lie very close to the heavy line. 
All the other points in the inset describe Type B transitions, 
 $\Psi(1,1) \to \Psi(1,3)$ (Point 1), $\Psi(1,3) \to \Psi(2,1)$ (Point 2),
 $\Psi(3,1) \to \Psi(2,3)$ (Point 4), $\Psi(2,3) \to \Psi(4,1)$ (Point 5),
 and $\Psi(5,1) \to \Psi(3,3)$ (Point 7). 
The Type B transitions tend to lie well below the heavy line,
 and these transitions involve a large change in the number of rows. 
The heavy line describes the situation when Type A transitions only occur,
 which can happen when the strength of the potential is increased.
 This is discussed in Section \ref{sec:pot}. 

\begin{figure}
%  \narrowtext
  \centerline{\epsfxsize=7.5cm
  \epsfbox{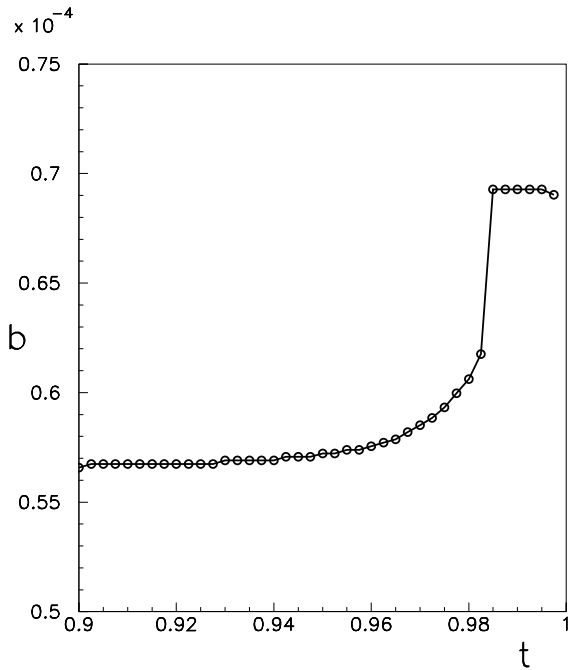} }
  \caption{The `melting' transition between states with $N_y=8$ and
   $N_y=9$ for $h=6.9\times10^{-4}$, $B_0 = 2\times 10^{-6}$}
  \label{fig:melt}
\end{figure}

% Melting Transition
The transitions can also be observed as the temperature is varied.
 For small $N_y$ these `melting' transitions only occur over a very small
 range of the applied field $h$. 
As the range of allowed values of $R$ increases the range of field over
 which the lattice can be seen to melt also increases. 
Fig. \ref{fig:melt} shows the transition between states $\Psi(8,\ge 1)$
 and $\Psi(9,\le 1)$. 
%insert figure
\vglue -13.0pt

%%%%%%%%
%%  London Theory
%%%%%%%%
\subsection{London Theory}

The LLL will not be strictly applicable in the limit of small $b$. The LLL
 approximation may be valid only in the region where
 $\pt \Delta b / \pt h < 0$ in Fig. \ref{fig:dbvsh}. 
By increasing $b_0$ it is possible to shift the maximum in $\Delta b(h)$
 to smaller values of $N_y$ but London Theory should be used to investigated
 the properties of the system with small values of $N_y$. 

The London free energy is 
\be
{\cal F}_{Lon} = { 1 \over 2 \mu_0 } \sum_{i\ts j} \int \int \hbox{d}
{\bf r}_i^\alpha V_{\alpha\beta}({\bf r}_i - {\bf r}_j)
 \hbox{d}{\bf r}_j^\beta
\label{londonfree}
\ee
where $V_{\alpha\beta}({\bf r)}$ is the potential defining the local
 magnetic induction $B({\bf r})^\beta = \sum_i \int \hbox{d}{\bf r}_i^\alpha
 V_{\alpha\beta}({\bf r} - {\bf r}_i)$, the parameters $\{\alpha,\beta\}$
 correspond to the $\{x,y,z\}$ components, and $\{i,j\}$ sum over the
 contributions from the all flux lines. 
We investigate an isotropic system, where the Fourier transform of the
 London potential for straight flux lines is $\widetilde{V}_{\alpha\beta}
({\bf k}) = \delta_{\alpha\beta} S({\bf k}) / (1 + \lambda^2 k^2 )$, 
 $\lambda$ being the London penetration depth. The cutoff function
 $S({\bf k}) = \exp(-\xi^2 k^2)$ removes the divergences within London
 Theory due to the absence of the flux line cores. 
The use of different cutoffs within London Theory has been discussed
 previously\cite{thompson}. 

We use similar geometries to the previous section and minimize
 ${\cal G}_{Lon} = {\cal F}_{Lon} - B\cdot H$ with respect to the positions
 of the flux lines. 
Again we assume the flux lines form a lattice commensurate with the principal
 region.

The low field behavior using the LLL was dominated by the minimum in
 $\beta_A(R)$. It might be thought that using London Theory could remove the
 multiplicity of transitions. 

The equilibrium lattice is always a centered rectangular structure, and the
 flux lines do not initially enter the principal region in a single
 straight row. 
This is different from a thin slab geometry, where straight rows of flux
 lines (rectangular symmetry) are frequently the equilibrium structure due
 to the surface barrier\cite{brongersma1,carneiro}. 

%Jumps London
%insert figure
\begin{figure}
%  \narrowtext
  \centerline{\epsfxsize=7.5cm
  \epsfbox{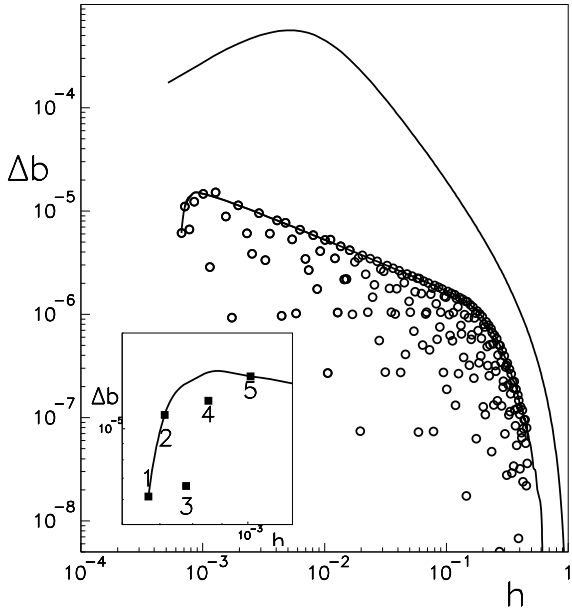} }
  \caption{The change in the magnetization as a function of the applied
 field $h$ for $L_y = 10\lambda$ and $\kappa = 50$. The upper heavy line
 shows the jumps associated with the LLL. }
  \label{fig:london}
\end{figure}

The flux lines enter as a centered rectangular structure with $N_y=1$, and
 there is a smooth crossover from $\Psi(1,< 1)$ to $\Psi(1,> 3)$. 
The first transition occurs between states $\Psi(1,>3)$ and $\Psi(2,<1)$,
 and the sequence of structures observed is then the same as the LLL limit. 
At each of the transitions there is an associated jump in the magnetic
 induction, but Fig. \ref{fig:london} shows that these are smaller than in
 the LLL limit. 
Again, during the Type B transitions (point 3, $\Psi(3,1) \to \Psi(2,3)$ and
 point 4, $\Psi(2,3) \to \Psi(3,1)$ ) the change in the magnetization is much
 smaller than during the Type A transitions (point 2, $\Psi(2,1)\to\Psi(3,1)$
 and point 5 $\Psi(4,1)\to\Psi(5,1)$ nearby. 
Therefore, despite the absence of the pronounced minimum in the Gibbs free
 energy the flux line lattice shows similar behavior to the LLL limit and
 the same sequence of states is observed.

\section{Non-zero potential}
\label{sec:pot}

In the previous section, the potential has not only been assumed to be
 periodic, {\it i.e. } $V(y) = V(y+L_y)$ but also to only restrict the
 choice of suitable flux line lattices. 
This may be either an extremely weak potential or one whose contribution
 to the total energy is zero. 
Now we investigate the equilibrium lattice when the periodic potential
 does contribute to the overall energy. 
We define the potential only in the principal region
 $V(y) = V_0 ( \cosh(a y) - 1 )$ where $-L_y/2 < y < L_y/2$ and assume the
 potential is periodic as described above. 
The potential is zero in the center of the principal region, but depending
 on the sign of $V_0$ will be either attract or repel flux lines form the
 edges of the principal region. 
This potential then favors states with $R=1/\sqrt{3}$ or those with
 $R=\sqrt{3}$. Assuming the potential is still weak, we include just the
 first order correction $\langle \Psi | V(y) | \Psi \rangle$ to the ground
 state energy. 

The potential favors the Type A transitions between states near the same
 minimum of $\beta_A(R)$ and the inclusion of this potential reduces the
 number of Type B transitions. 
As $V_0$ is increased it is then possible to remove transitions between the
 two minima in $\beta_A(R)$ and for transitions to occur between states with
 $R\sim 1/\sqrt{3}$ or between states with $R\sim \sqrt{3}$. 
In the figure analogous to Fig. \ref{fig:scheme} the allowed values of $R$
 would be one shaded band or the other and there would be no transitions
 between them.

The repulsive (attractive) potential ensures that the lattice chooses
 alignments aligned parallel (perpendicular) to the periodic direction, i.e.
 the $\hat{\bf y}$-direction. 
The jumps in the magnetic induction follow the heavy curve in
 Fig. \ref{fig:dbvsh}. This behavior also occurs in the London limit, where
 the inclusion of a potential again favors the existence of states
 permanently aligned either parallel or perpendicular to the y-axis.

%insert figure
\begin{figure}
%  \narrowtext
  \centerline{\epsfxsize=7.5cm
  \epsfbox{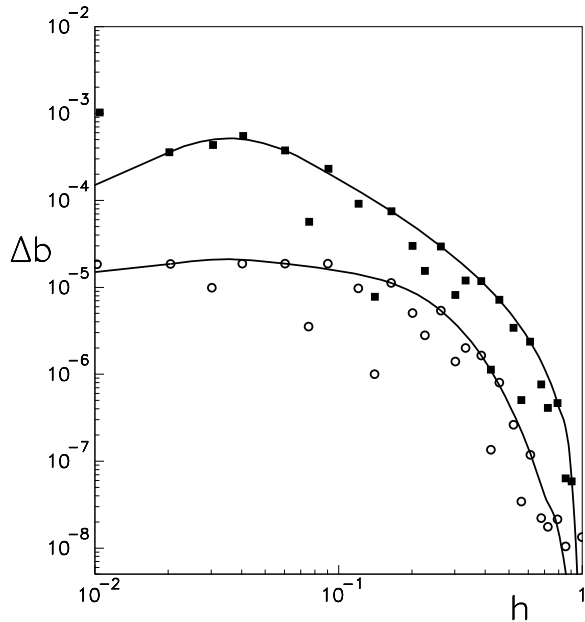} }
  \caption{The changes in the magnetization for $L_y = \lambda$, showing
 both the LLL limit (filled squares) and the London limit (open circles). }
  \label{fig:dbvsh1}
\end{figure}
\begin{figure}
%  \narrowtext
  \centerline{\epsfxsize=7.5cm
  \epsfbox{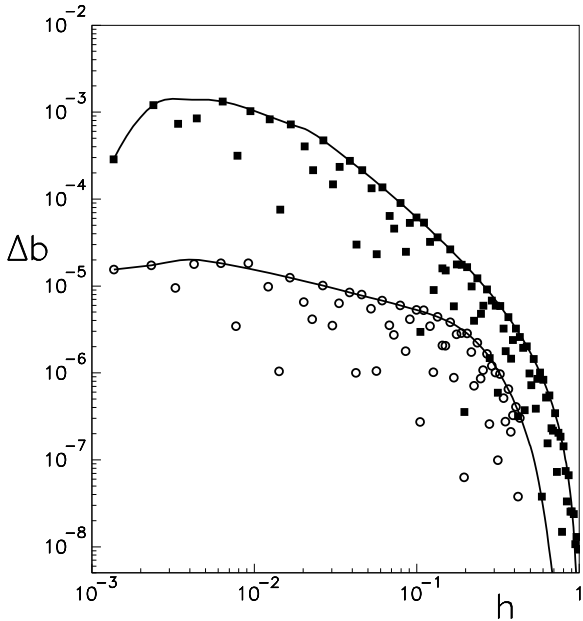} }
  \caption{The changes in the magnetization for
 $L_y = \protect{\sqrt{10}}\lambda$ showing both the LLL limit
 (filled squares) and the London limit (open circles). }
  \label{fig:dbvshs10}
\end{figure}

The behavior in Fig. \ref{fig:dbvsh} for $L_y/\lambda = 10$ is controlled
 by $B_0$. Reducing the width of the periodic potential at fixed penetration
 depth $\lambda$, the maximum in $\Delta b(h)$ moves slightly, to
 smaller $N_y$. 
Fig. \ref{fig:dbvsh1} and Fig. \ref{fig:dbvshs10} show the jumps in the
 magnetization for different widths. The LLL (filled squares) and the
 London limit(open circles) are shown together and can be compared with
 Fig. \ref{fig:dbvsh} and Fig. \ref{fig:london}. 
The heavy lines indicate the transitions that occur in the strong potential
 limit, and all transitions disappear in the limit $h\to 1$. The behavior
 is very similar for the three different widths shown, but increasing the
 width increases the number of transitions observed. 
%insert figure

Theunissen et al\cite{kes} have investigated the properties of a
 NbN/Nb$_3$Ge bilayer under the influence of an applied current and magnetic
 field. 
A regular array of channels was etched through the thin NbN top layer into
 the Nb$_3$Ge. 
The NbN layer, with its much higher critical current than the Nb$_3$Ge
 layer effectively acted as a pinning center for the flux lattice. 
By measuring the shearing force needed for the flux lines to move along the
 channels, the shear modulus of the flux line lattice could be measured. 
The shear modulus shows a characteristic behavior
 $c_{66}\propto b ( 1 - b^2 )$ as the field is increased. In the London limit
 $c_{66}\propto b$ while in the LLL limit $c_{66}\propto (1-b^2)$. 
Superimposed on this functional form was an oscillatory function that
 reflected the finite nature of the channel width. 

The shear modulus $c_{66} \propto \pt^2 F_{vol} / \pt \alpha^2$ where
 $\alpha $ is the shear angle. In the LLL limit, $F_{vol} \propto 1/\beta_A$. 
The Abrikosov parameter can be written as a sum over all reciprocal lattice
 vectors ${\bf Q}$, $\beta_A = \sum_{{\bf Q}} \exp - Q^2/2U^2$, where $U$
 is the inverse magnetic length $U^2 = 2\pi / A_{unit}$, $A_{unit}$ being
 the area of the unit cell. This allows direct calculation of the shear
 modulus for all lattices. 

%c_66
%insert figure
\begin{figure}
%  \narrowtext
  \centerline{\epsfxsize=7.5cm
  \epsfbox{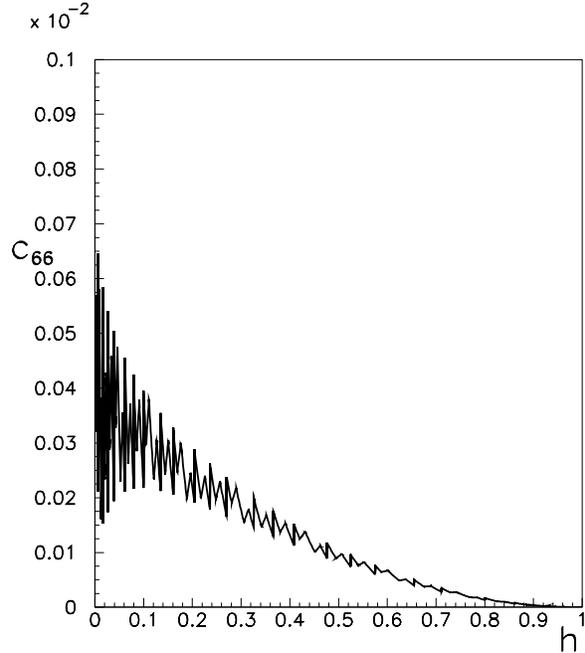} }
  \caption{The shear modulus for the commensurate flux line lattice with no
 added potential, showing the oscillatory nature as a function of the
 magnetic induction, for $L_y = \protect{\sqrt{10}} \lambda$ and
 $\kappa=50$. }
  \label{fig:c66}
\end{figure}

The modulus changes dramatically as the lattice reconstructs. 
Fig. \ref{fig:c66} shows $c_{66}$ in the LLL , where the oscillations
 characterize the rearrangement of the flux line lattice within the channel. 
This shows more oscillations than the experiments of Theunissen
 et al\cite{kes}, and the introduction of the (attractive) potential models
 the experiments more closely. 
The size and positions of the transitions that Theunissen et al.\cite{kes}
 observed indicate transitions only with $\Delta N_y = +1 $ and for
 $R\sim \sqrt{3}$. Therefore, the effects of the interaction of the flux
 lines with the underlying periodic potential are likely to be important,
 although the interaction with impurities may hinder extensive
 reconstructions with large changes in $N_y$.

\section{conclusion}

For an infinite system it is well known the flux lines form a periodic
 lattice with hexagonal symmetry. Once there is competition with an
 underlying potential this is not necessarily the situation. 
The flux lines still favors lattices close to this ideal lattice, and
 transitions can be seen between different structures. 
Using both London Theory and the LLL approximation we have investigated some
 of the properties associated with these structures. As the applied field is
 increased there are notable jumps in the magnetic induction, and hence in
 the magnetization and critical current. These transitions occur in two ways. 
The Type A transitions just involve an increase in the number of rows of flux
 lines in the principal region. Type B transitions occur between states
 aligned parallel and perpendicular to the periodic direction. 
The different alignments of the lattice correspond to the two competing
 minima in $\beta_A(R)$. Increasing the strength of the periodic potential
 ensures one of these alignments is favored over the other and eventually
 only the Type A transitions increasing the number of rows occur. 
For $V_0$ less(greater) than zero, the potential favors the flux lines being
 at  the edges(center) and the flux line lattice is aligned
 parallel(perpendicular) to the periodic direction. 

The transitions are most easily seen as the applied field is increased.
 However they also occur as the temperature is varied and the lattice can
 appear to melt as the temperature is increased. Also associated with these
 transitions are changes in the physical properties of the flux line lattice,
 such as the shear modulus. 
The matching fields and the nature of the $c_{66}$ indicate the experiments
 of Theunissen et al.\cite{kes} are best modeled by an attractive potential,
 which pulls the flux lines to the edges of the region. 

\acknowledgements{It is a pleasure to thank S. Bending for stimulating
 conversations. This work was supported by EPSRC through grant GR/J60681.}


\begin{references}

\bibitem{martinolli} P. Martinolli, M. Nsabimana, G.A. Racine and H. Beck,
 Helv. Phys. Acta {\bf 55}, 655 (1982).

\bibitem{raffy} H. Raffy, E. Guyon and J.C. Renard,
 Solid State Comm. {\bf 14}, 431 (1974).

\bibitem{metlushko} V.V.~Metlushko, M.~Baert, R.~Jonckeere, V.V.~Moshchalkov
 and  Y. Bruynseraede, Solid State Comm. {\bf 91}, 331 (1994).

\bibitem{cooley} L.D. Cooley, P.J. Lee and D.C. Larbalestier,
 Appl. Phys. Lett. {\bf 64}, 1298 (1994).

\bibitem{oboznov} V.A. Oboznov and A.V. Ustinov,
 Phys. Lett. A {\bf 139}, 481 (1989).

\bibitem{brongersma1} S.H. Brongersma, E. Verweij, N.J. Koeman,
 D.G. de Groot and R.Griessen, Phys. Rev. Lett. {\bf 71}, 2319 (1993).

\bibitem{bolech} C. Bolech, G.C. Buscaglia and A. Lopez,
 Phys. Rev. B {\bf 52}, 15719 (1995).

\bibitem{ziese} M. Ziese, P. Esquinazi, P. Wagner, H. Adrian,
 S.H. Brongersma and R. Griessen, Phys. Rev. B {\bf 53}, 8658 (1996).

\bibitem{kes} M.H. Theunissen, E. Van der Drift and P.H. Kes, 
 Phys. Rev. Lett. {\bf 77}, 169 (1996).

\bibitem{fischer} O. Brunner, J.M. Triscone, L. Antognazza, L. Meiville,
 M.G. Karkut and O. Fischer, J. Less Comm. Metals {\bf 164}, 1186 (1990).

\bibitem{kleiner} 
W.H. Kleiner, L.M. Roth, and S.H. Autler,
 Phys. Rev. {\bf A1226}, 133 (1964).

\bibitem{thompson} A.M.Thompson and M.A.Moore, 
 Phys. Rev. B {\bf 55}, 3856 (1997).

\bibitem{carneiro} G. Carneiro, preprint cond-mat/9707248.

\end{references}
\end{document}